\documentstyle[psfig]{ioplppt} 
\begin{document}
\jl{1}

\title{Stationary Self-Organized Fractal Structures in an Open,
Dissipative Electrical System}  
\letter
\author{Marco Marani\dag\ftnote{3}{To whom correspondence should be addressed.}, Jayanth R. Banavar\ddag, Guido Caldarelli\P,
 Amos Maritan$\Vert$, and Andrea Rinaldo\dag}

\address{\dag Dipartimento Di Ingegneria Idraulica, Marittima e Geotecnica e 
 Centro Internazionale di Idrologia "Dino Tonini", Universit\`a di Padova, 
via Loredan 20, I-35131 Padova, Italy }
\address{\ddag Department of Physics and Center for Materials Physics, 
The Pennsylvania State University, 
104 Davey Laboratory University Park, PA  16802 USA}
\address{\P TCM Cavendish Laboratory, Madingley Road, CB3 OHE, Cambridge, UK}
\address{$\Vert$ International School for Advanced Studies, 
I-34014 Grignano di Trieste, 
Istituto Nazionale di Fisica della Materia and sezione INFN di Trieste, Italy}

\renewcommand{\baselinestretch}{2}

\begin{abstract}
We study the stationary state of a Poisson problem for  
a system of $N$ perfectly conducting metal balls driven by electric 
forces to move within a medium of 
very low electrical conductivity onto which  charges are sprayed from
outside. 
When grounded at a confining boundary, the system of metal balls 
is experimentally known to self-organize  
into stable fractal aggregates. 
We simulate the dynamical 
conditions leading to the formation of such aggregated patterns
and analyze 
the fractal properties.
 From our results and those obtained for steady-state systems 
that  obey minimum total energy 
dissipation (and potential energy of the system as a whole),  
we suggest a possible dynamical rule for the emergence of scale-free
structures   
in nature.
\end{abstract}
\vspace{-1cm}

\maketitle

\section{Introduction}

The key ingredients for the ubiquitous appearance of natural phenomena
lacking 
a spatial and/or temporal scale remain a mystery.
Self-organized criticality\cite{Bak1, Bak3, Bak7} 
 represents one basic attempt 
towards the elucidation of a general mechanism responsible for scale and 
time invariance in a variety of phenomena in diverse fields including
physics,  biology, 
geophysics and economics.  An alternative view  has been postulated 
on the basis of fractal forms occurring in natural river basins and 
in domain walls of random ferromagnets\cite{ocn1,ocn3,book,ivsla,ferr}.  It has
been speculated 
that chance and necessity, defined by an  
imperfect strive 
for global optimality of open, dissipative systems could be
ingredients for some forms of natural fractal growth.   
In this paper,  we study the scale-free stationary states resulting 
from the 
dynamics of an open electric system.  The steady state configuration
corresponding to minimum total potential energy is a periodic non-fractal
structure.  
Our findings suggest a rule for the emergence of fracatality that encompasses 
dynamical problems with a global constraint.

We propose a model for the spontaneous aggregation of metal balls 
floating in a viscous liquid having a high dielectric constant.
Electric charges sprayed on to the poorly conducting 
still fluid are transported to a grounded confining ring
by the metal balls. Such aggregation has been observed\cite{merte1, merte2}
in cylindrical cells 
made up by a layer of castor 
oil (because of its high dielectric constant) within an acrylic 
dish confined by a layer of air above. The inner perimeter of the 
dish was equipped with 
a grounded metal ring. A strong potential between a metallic tip and this 
ring was fixed and  charges were sprayed 
quasi homogeneously upon the oil surface. Inside the dish there were 
metallic bearing balls, which  spontaneously 
arrange themselves under the influence of  electric forces that lead to the
transportation of charges to the metal ring. It is relevant to our
computations 
that the reference experiment was repeated 
several times for different 
numbers of balls and, at stationary conditions, the aggregated 
structures showed fractal properties\cite{merte1, merte2}.  

\section{Description of the model}
Our model consists of $N$ particles (the metal balls of experiment
\cite{merte1, merte2}) in a domain
 $\Gamma$ with $N_\Gamma$ ($\gg N$) sites in a $2-d$ square lattice.
 The particles are charged (due to constant injection of charges from
outside) and are subject to
 the force due to the electric potential $\varphi({\bf x})$ which obeys the
Poisson equation

\begin{equation}
\nabla^2 \varphi=-S_0/\sigma_0=\rho({\bf x})
\label{poisson}
\end{equation}

where $S_0$ is the source term, $\sigma_0$ is the specific conductivity of
the medium and $\rho$ is
 the charge density. Eq. (\ref{poisson}) follows from charge conservation
and Ohm's law \cite{merte1, merte2}.
The charge density is assumed to be constant in space (i.e. perfectly
homogenous spraying
of charges) and, by a suitable rescaling of the field $\varphi$, 
its value may be assumed equal to 1 and is immaterial as long as it is not
zero.
The Laplacian in eq. (\ref{poisson}) is a lattice discretization of the
corresponding continuum
operator and it is computed assuming a eight-neighbour scheme (four
neighbours 
along the vertical and horizontal directions and four along the diagonals).

Poisson's eq. (\ref{poisson}) is solved \cite{numerics} with the boundary
condition
 $\varphi(\partial \Gamma')=0$, where $\partial \Gamma'$ includes the
boundary, $\partial \Gamma$,
 of $\Gamma$ and all of the sites in $\Gamma$ which are occupied by a
particle connected to
 $\partial \Gamma$ through other particles. The particles in $\partial
\Gamma$ are immobile.
This assumption considerably reduces the more complex system studied in
\cite{merte1, merte2}, 
and is qualitatively akin to the experimental conditions where the
difference in 
electrical conductivity between the metal balls and the embedding 
castor oil is quite large.

Initially the $N$ particles are placed at random in $\Gamma$ and $\partial
\Gamma'= \partial \Gamma$,
 i.e. no particles are in contact with the boundary. The potential
$\varphi({\bf x})$ is computed at all
$N_\Gamma$ sites by solving eq. (\ref{poisson}) with b.c.
$\varphi(\partial \Gamma)=0$.
Particles not belonging to $\partial \Gamma'$ are considered immaterial to
the evaluation of
$\varphi$ because of their assumed very large conductivity.
The gradient $\nabla_\mu \varphi({\bf x})$ in all eight directions
($\mu=1,\ldots,8$), at all sites
 ${\bf x}$ ($\not\in \partial \Gamma'$) is now computed. The largest value
 $|\nabla_{\hat{\mu}} \varphi ({\bf x})|$ is then found and the
corresponding particle is moved in the
 direction $\hat \mu$ (in case such a move caused double occupancy of a
cell the site
 with the next largest value of $|\nabla \varphi ({\bf x})|$ is selected
instead; and so on).  This is then repeated until
a particle becomes a neighbour of the boundary $\partial \Gamma'$ and thus
a part of the boundary.
Whenever this occurs the potential $\varphi ({\bf x})$ must be recomputed
due to the change in boundary conditions.
This process is repeated until all the particles eventually become part of
the boundary.
 
Other dynamics may be used in which 
all the particles are moved simultaneously in the direction of the maximum
 gradient. The stationary configurations that emerge are qualitatively 
 similar  to those obtained through the procedure described here.

It is crucial that at stationary conditions for the arbitrary $i$-th metal
ball 
located at ${\bf x} = {\bf q}_i$, (i.e. when ${\bf {\dot q}}_i 
= {\bf {\ddot q}}_i = 0$) the total potential energy 
${\cal W} \propto \int_{\Gamma} \varphi d{\bf x}$ 
is at a minimum with respect of the arrangement of the
metal balls 
(i.e. $\nabla_{{\bf q}_i} {\cal W} =0$) \cite{merte1, merte2}. Furthermore, 
the system configuration 
that minimizes ${\cal W}$ also exactly 
minimizes the total 
energy dissipation $P \propto \int_{\Gamma} |\nabla \varphi|^2 d{\bf x}$
\cite{merte1, merte2}. 
The above framework relates directly to other contexts, where scale
invariance was argued 
to arise as a consequence of frustrated optimality \cite{book}. 

The computation of the potential $\varphi({\bf x})$ allows the
determination of 
the total electric resistance, say $R$ which, in the  experiment, 
had been shown to scale with the number $N$ of metal balls as
$ R \propto N^{-\xi}$.   
Box counting and cluster 
dimensions of the sets formed by the connected aggregates were also
computed and  
the scaling of $N_1(l)$, the number of boxes of size  
$l$ making up the structure, and of $N_2(r)$, the number of balls inside 
a radius $r$ was studied.

Technically, the solution of a Poisson equation at each 
time step 
implies a considerable numerical burden and an efficient 
numerical scheme had to be employed. Finite difference 
approaches require the solution of a $L^2 \times L^2$ linear 
system, where $L$ is the size of the lattice. The use of the 
accelerated conjugate gradient technique\cite{numerics}, though very
efficient 
when a single solution is required, 
does not take advantage of 
the fact that two successive fields only have modest 
differences due to the local variation of the boundary conditions. 
A combined use of a relaxation technique and an initial 
direct solution proved computationally more advantageous, 
because it bypasses the slow relaxations   
when first solving the equations, 
and updates, given the modified boundary conditions, the 
configuration of $\varphi({\bf x})$ and quickly converges.
Note that 
the improvement in efficiency was especially critical when 
exploring a substantial number of possible 
perturbations of a given arrangement obtained from the dynamics to test its 
optimality.

\section{Numerical experiments}

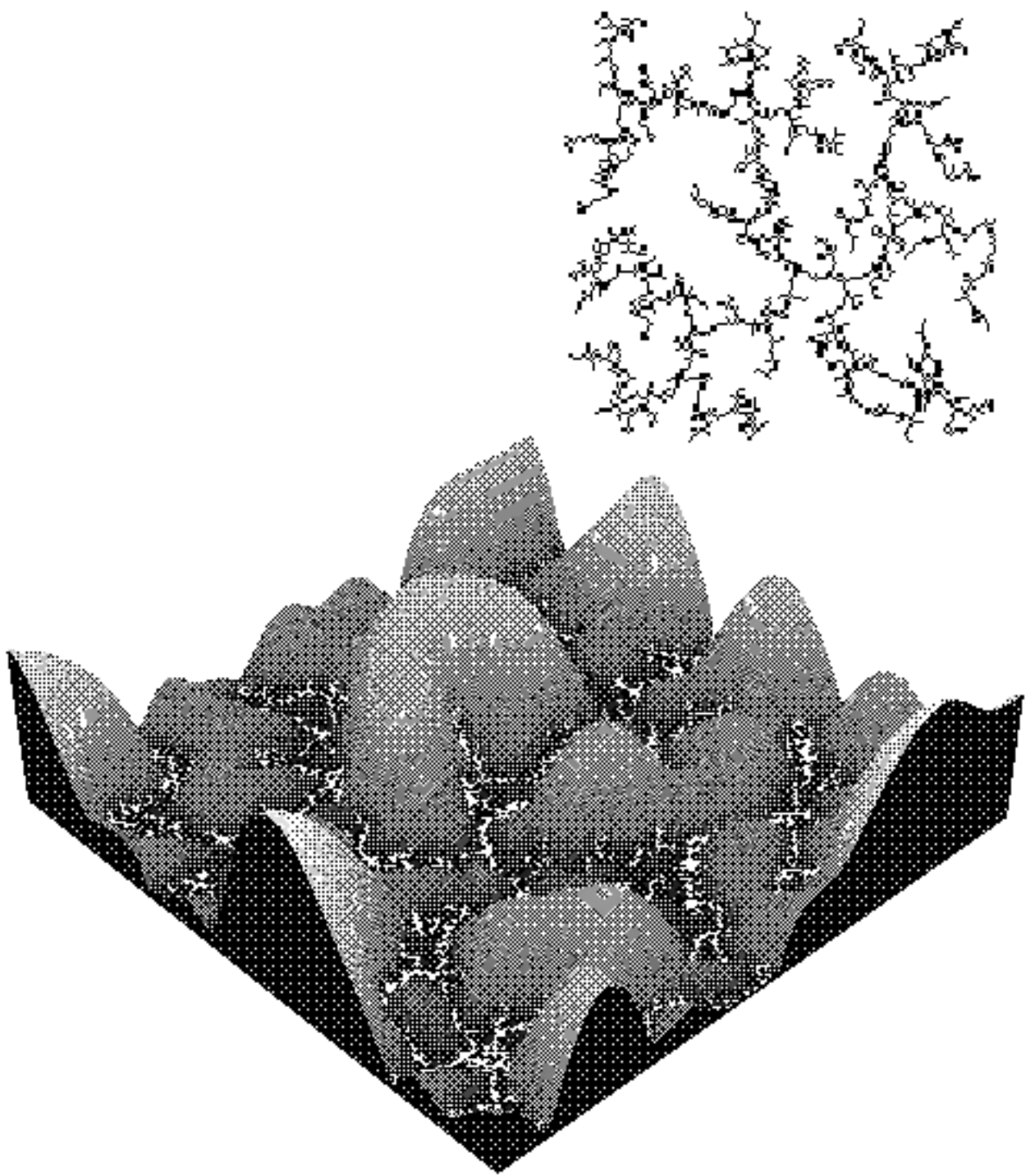
\begin{figure}
\caption[fig1]{A realization of the potential field with $N=2048$ metal
balls within 
a $128 \times 128$ grid grounded at the central site ${\bf P}$ (i.e.
$\varphi({\bf P}) =0$). 
In the inset we show the planar aggregation structure of the grounded 
metal balls. The total potential energy is ${\cal W} = 14.0$ (in arbitrary
units).}
\label{f1}
\end{figure}

We have run several simulations of the dynamics of the system where
initially the 
$N$ balls 
are randomly distributed over the domain $\Gamma$,   
and have obtained consistent fractal patterns for the aggregate arrangement 
of the grounded elements. Figure \ref{f1} shows one 
realization of the final potential field, in 
this case related to a square domain grounded at its center (i.e. $\partial
\Gamma$ 
consists of a single site) where $N=2048$ aggregated 
metal balls are employed. We have used these particular boundary
conditions, at no 
cost of generality, to 
compare the final structure with an exact optimal arrangement described
below. 
The shades of gray in Figure 1 are related to the 
local value of the potential 
$\varphi({\bf x})$. We have examined in detail  
the statistics of the resulting aggregates at different densities
$N/N_\Gamma$.  
Figure 2 shows sample results of finite-size scaling analyses 
related to the estimation of box-counting dimensions\cite{man} and the mass 
dimension $D_m$\cite{feder}. We have also computed the 
scaling exponent $\xi$ of total resistance, $R \propto
N^{-\xi}$\cite{merte1, merte2}.  
Nevertheless it is not possible to compare scaling exponents from our 
numerical simulations and observations. In fact, the dynamical model is 
 simple and does not
capture all the experimental details.  We have made simplifying assumptions
for
computational convenience, e.g. 
infinite electric conductivity of the metal
balls, 
zero mechanical inertia of the metal balls and  infinite cohesive forces
among the grounded balls.

In the case of Figure \ref{f1} ($N=2048$) for a square grid 
of size $L=128$, the 
box-counting exponent is $D = 1.30$, the mass dimension 
$D_m = 1.63$  (Figure \ref{f2}) and the scaling  
exponents of resistances is $\xi = 2.00$. In this case  
the total potential energy is ${\cal W} = 14.9$ (in arbitrary units) and in
all 
realizations with the same  boundary conditions and  number $N$ the
final value 
of ${\cal W}$ is within a few units. The 
statistics of the algebraic laws are robust. We have also tested  
circular and square domains grounded at the outer boundaries -- one obtains
DLA-like dendrites issuing from the center with similar results.  
Note that in all cases the  
dynamics is not stochastic, differently from the cases of DLA and
DBM\cite{Sander1, Sander2}.

As one expects, the global minimum is very unlikely 
to be accessible by the dynamics. In fact 
the structure having minimum total potential ${\cal W}$, which we have 
found exactly for regular geometries (It can be shown that a regular grid of $N$ connected 
grounded points is the global minimum of total potential ${\cal W}$.   
The proof proceeds from a one dimensional situation in which
the arrangement of arbitrary $N$ grounded points within a field obeying 
$\varphi" = 1$ pinched to zero at the boundaries is a regularly spaced 
array of zeroes. The extension to two dimensions comes from 
separation of variables of the 2D Poisson equation and the requirement 
of connectivity of the grounded structure), is characterized 
by great regularity and substantially different scaling 
exponents (Figure 3). 
In the case of the square domain occupied by 
$N=2048$ connected sites of zero potential, a regular grounded lattice 
yields the optimal energy ${\cal W}=5.60$, and a box-counting and mass 
exponents $D=D_m=2.00$, outside the error bars of dynamically accessible
states.

The resulting structures, in all cases, do indeed correspond to local
minima of 
the total potential energy. 
Hence, in the exactly known case of a regular arrangement and in analogy with other 
physical processes\cite{book,ivsla,ferr,science}, imperfect optimal search 
procedures yield 
local optima showing scaling properties quite different from those of 
the true ground state. 
In the general case, the exact properties of the ground state are not known, 
and thus we have used   
refined annealing procedures\cite{fo171,fo172} starting  from the
self-organized  
aggregates to search for optimal aggregation. This procedure
 uses a randomized approach in the exploration of the values of the 
 objective function in order to avoid being trapped in local minima and it is
 known to ensure achievement of values close to the absolute minimum.
 The arrangements obtained through this optimal search also show 
 significant departures from the features of the initial structure.

\section{Conclusions}
The numerical experiments described before  
seem to indicate that the type of optimization that is dynamically feasible
in nature 
is rather myopic, that is, willing to accept changes 
only if their impact is 
favourable in the immediate  
rather than in the long run. 
Whether or not dynamical rules necessarily obey a variational 
principle, and thus whether fractal forms in nature are or are not
metastable states of a dynamical search for optimality, our results suggest  
that one ingredient for the emergence of fractality
is the existence of a set of 
stationary or recursive states satisfying a global constraint. 
In our  case the constraint is the dissipation of the injected charge
through the boundaries. 
It is thus possible that, in some cases, local interactions may     
yield a globally felt constraint through boundary conditions. 
The system studied here is quite analogous to glasses.  There,
the ground state is known to be a crystal but on rapid quenching of
a liquid, the dynamics
inhibits crystallization and a glassy structure
(that is not fractal) is obtained.  The key difference 
is that here the boundaries 
impose a global constraint that result in the metastable states
becoming scale-free over a certain range of length scales.

\vskip 1.0cm
{\footnotesize

We are indebted to Ignacio Rodriguez-Iturbe, Riccardo Rigon and Marek
Cieplak for stimulating
discussions.
This work was supported by grants from  NATO, CNR-GNDCI 
(Progetti MIEP-METEO, Linea 1) and  MURST 40 \%.}

\newpage

\begin{figure}
\caption[fig2]{(a) Box-counting scaling relationships, $N_1(l) \propto l^{-D}$, for a varying
number of balls 
within a $128 \times 128$ domain. 
Notice the consistency in the slopes; (b) the computation of the mass 
dimension $D_m$ requires the evaluation of the exponent of the scaling
relationship 
$N_2(r) \propto r^{D_m}$ where $N_2(r)$ is the number of balls falling within a
circle of radius 
$r$ seeded at ${\bf P}$. The computation was carried out over several
realizations obtained 
for $2048$ metal balls on a $128 \times 128$ gridded domain. The departure
from a power law for large 
$r$ is due to finite-size effects. In all plots larger symbols indicate
the ensemble 
mean while dots indicate single realizations.}
\label{f2}
\end{figure}

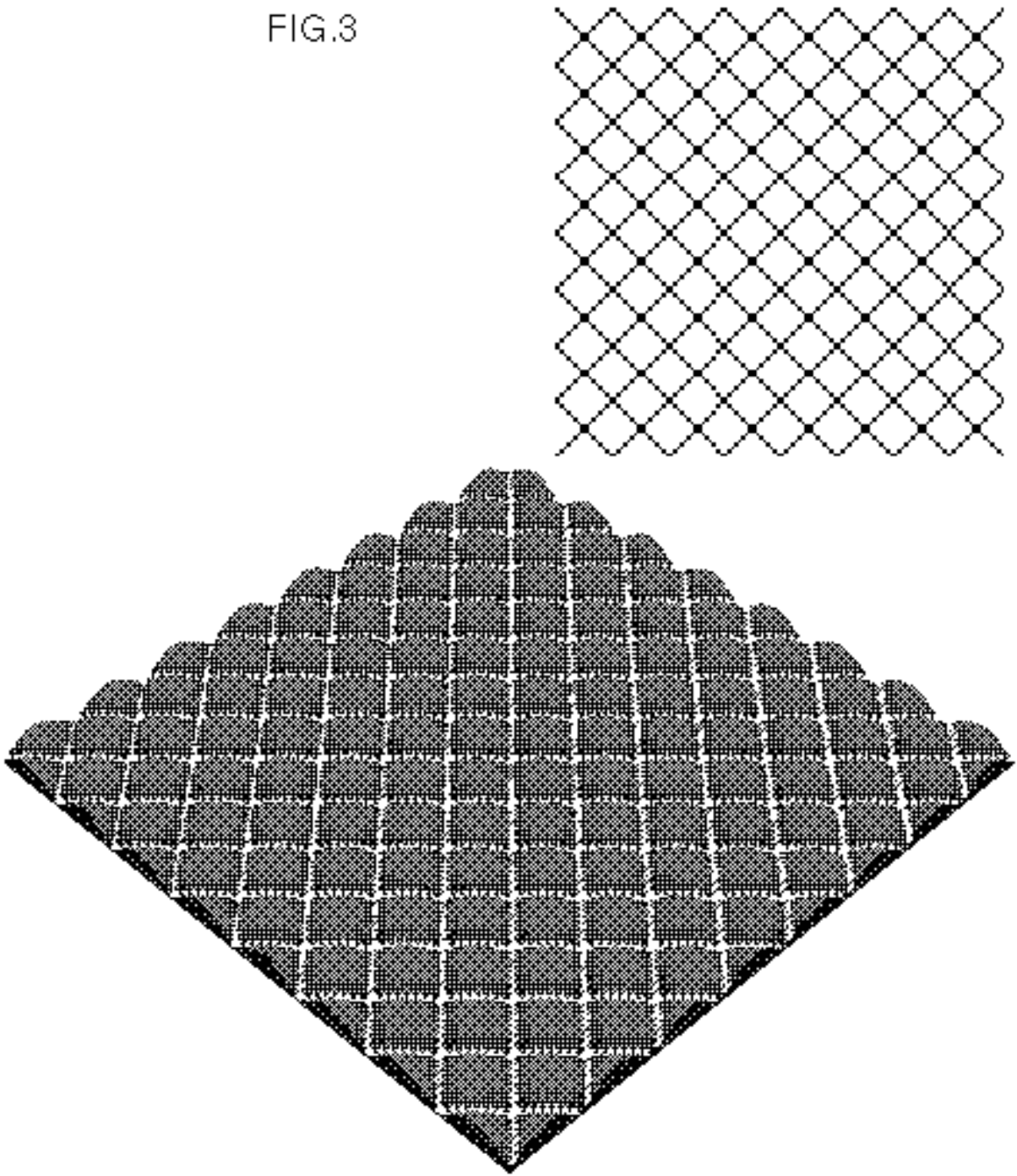
\begin{figure}
\caption[fig3]{Potential field arising from a regular, grid-like
arrangement of 
$N=2048$ metal balls within a center-grounded square $128 \times 128$
domain.  
${\cal W} = 5.6$ (in the same units employed for the system in Figure 1) 
and is substantially lower than the
values obtained by 
spontaneous self-organization and by imperfect optimal search techniques. }
\label{f3}
\end{figure}


\Bibliography{99}

\bibitem{Bak1} Bak P., C. Tang and K. Wiesenfeld, 1987, 
Phys. Rev. Lett., {\bf 59}, 381; 

\bibitem{Bak3} Bak P., K. Chen and M. Creutz, 1989, Nature, {\bf 342}, 780;

\bibitem{Bak7} Bak, P.,  1996, {\it How nature works} (Springer-Verlag, New York).

\bibitem{ocn1} Rodriguez - Iturbe I.,
A. Rinaldo, R. Rigon, R.L. Bras and E. Ijjasz - Vasquez, 1992,
Water Resour. Res., {\bf 28}, 1095;

\bibitem{ocn3} Rinaldo A., I. Rodriguez-Iturbe, R. Rigon, R.L. Bras, 
E. Ijjasz-Vasquez and A. Marani,  1992,
Water Resour. Res., {\bf 28}, 2183;

\bibitem{book} Rodriguez-Iturbe  I. and A. Rinaldo, 1997, {\it Fractal 
River Basins: Chance and Self-Organization} (Cambridge Univ. Press, 
New York);

\bibitem{ivsla} Rinaldo A., A. Maritan, F. Colaiori, A. Flammini, R. Rigon, 
M.R. Swift, J.R. Banavar and I. Rodriguez-Iturbe, 1997, Atti dell'Istituto 
Veneto di Scienze, Lettere ed Arti, {\bf 26}, 224;

\bibitem{ferr} Swift M. R., A. Maritan and J. R. Banavar, 1997, {\it Phys. Rev.
Lett.}, {\bf 77}, 5288;

\bibitem{merte1} Mert\'e  B. et al., 1989, Helvetica Physica Acta, {\bf 62}, 
294;

\bibitem{merte2} Hadwich G., B. Mert\'e, E. Luscher, 1990, Herbsttagung 
der SPG/SSP, {\bf 63}, 487;

\bibitem{numerics} Press W. H., S.A. Teukolski, W.T. Vetterly, and B.P.
Flannery, 1992, {\it Numerical Recipes} (Cambridge Univ. Press, New York);

\bibitem{man} Mandelbrot B.B., 1983, 
{\it The Fractal Geometry of Nature} (Freeman, New York);

\bibitem{feder} Feder J., 1988, {\it Fractals}, Plenum, New York;

\bibitem{Sander1} Witten T.A. and L.M. Sander, 1983, {\it J. Phys. A}, {\bf 16},
3365; 

\bibitem{Sander2} Niemeyer L., L. Pietronero and H.J. Wiesmann, 1984, {\it Phys. Rev. Lett.}, {\bf
52}, 1033;

\bibitem{science} Maritan A., F. Colaiori, A. Flammini, M. Cieplak, J. R.   
Banavar, 1996, {\it Science} {\bf 272}, 984; 

\bibitem{fo171} Metropolis N., M. Rosenbluth, M. Teller and E. Teller, 1953,
J. Chem. Phys., {\bf 21}, 1087;

\bibitem{fo172} S. Kirkpatrick, G. D. Gelatt \& M. P. Vecchi, 1993, {\it Science}
 {\bf 220}, 671;

\endbib
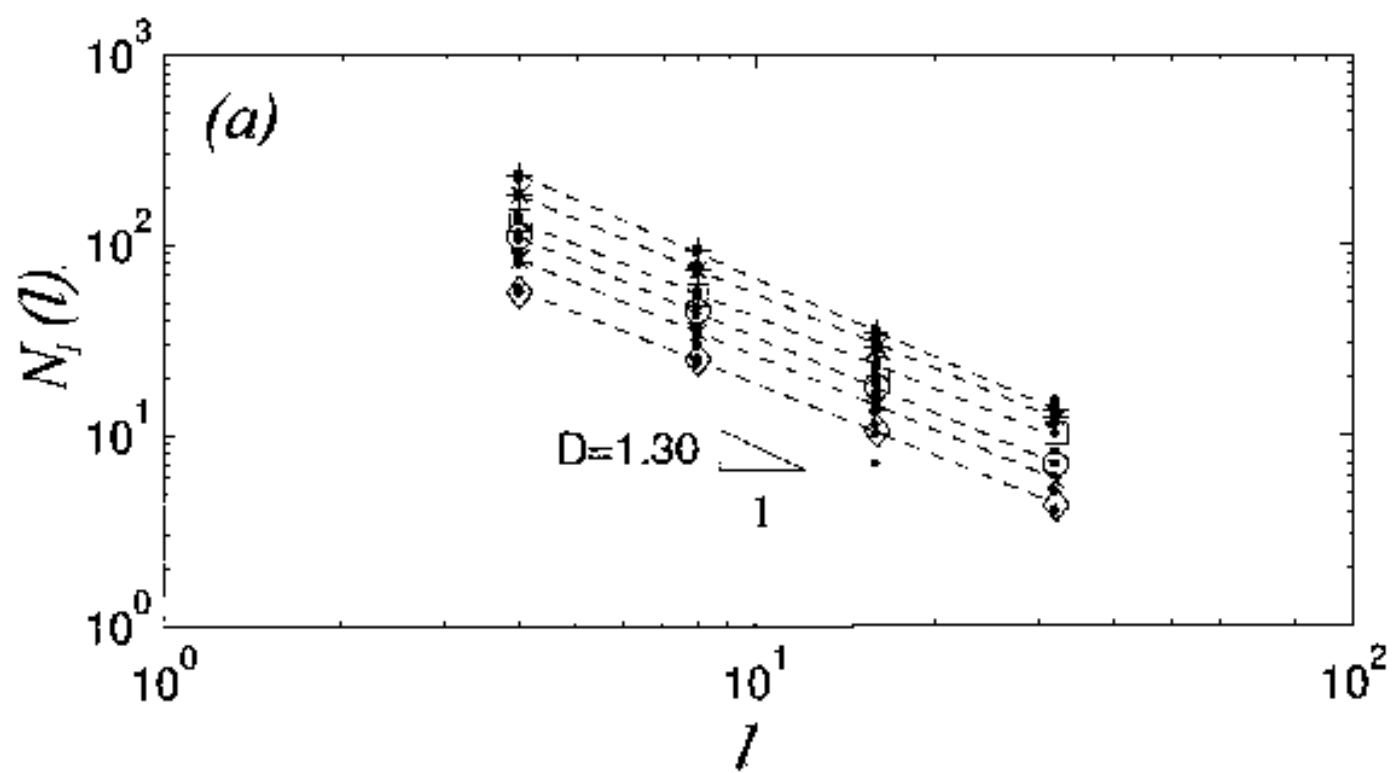
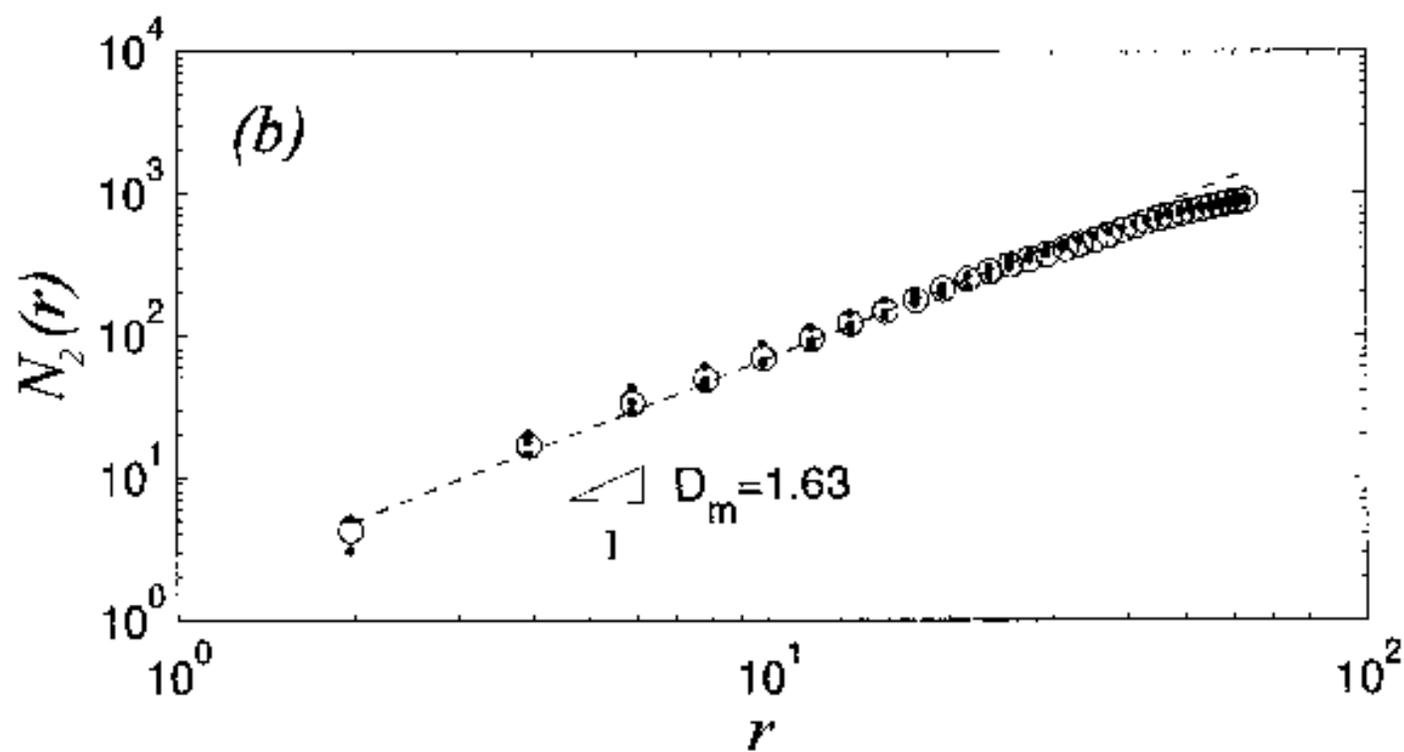
\end{document}